\documentclass{acm_proc_article-sp}

\usepackage{algorithm}
\usepackage[noend]{algpseudocode}
\usepackage{amssymb}
\usepackage{amsmath}
\usepackage{booktabs}
\usepackage[font=bf,labelfont=bf]{caption}
\DeclareCaptionType{copyrightbox}
\usepackage{color}
\usepackage{dashrule}
\usepackage{enumerate}
\usepackage{graphicx}
\usepackage{mathabx}
\usepackage{mdwlist}
\usepackage{microtype}
\usepackage{paralist}
\usepackage{setspace}
\usepackage{tabularx}
\usepackage{textcomp}
\usepackage[normalem]{ulem}
\usepackage{xcolor}

\renewcommand{\S}{\mathbf{S}}

\newcommand{\squishlist}{
 \begin{list}{$\bullet$}
  { \setlength{\itemsep}{0pt}
     \setlength{\parsep}{3pt}
     \setlength{\topsep}{3pt}
     \setlength{\partopsep}{0pt}
     \setlength{\leftmargin}{1.5em}
     \setlength{\labelwidth}{1em}
     \setlength{\labelsep}{0.5em} } }

\newcommand{\squishlisttwo}{
 \begin{list}{$\bullet$}
  { \setlength{\itemsep}{0pt}
     \setlength{\parsep}{0pt}
    \setlength{\topsep}{0pt}
    \setlength{\partopsep}{0pt}
    \setlength{\leftmargin}{2em}
    \setlength{\labelwidth}{1.5em}
    \setlength{\labelsep}{0.5em} } }

\newcommand{\squishend}{
  \end{list}  }

\allowdisplaybreaks

\newfont{\mycrnotice}{ptmr8t at 7pt}
\newfont{\myconfname}{ptmri8t at 7pt}
\let\crnotice\mycrnotice%
\let\confname\myconfname%

\toappear{\the\boilerplate\par
{\confname{\the\conf}} \the\confinfo\par \the\copyrightetc}

\permission{Copyright is held by the author/owner(s).}
\conferenceinfo{SIGIR'14 Workshop on Gathering Efficient Assessments of Relevance (GEAR'14),}{\\July 11, 2014, Gold Coast, Queensland, Australia.}
\copyrightetc{}

\begin{document}

\title{The Anatomy of Relevance}
\subtitle{Topical, Snippet and Perceived Relevance in Search Result Evaluation}

\author{
        Aleksandr Chuklin\titlenote{Now at Google Switzerland.}\qquad Maarten de Rijke\\[1.2ex]
        \affaddr{University of Amsterdam, Amsterdam, The Netherlands}\\
        \email{a.chuklin, derijke@uva.nl}
}

\maketitle

\begin{abstract}
    Currently, the quality of a search engine is often determined using so-called topical
    relevance, i.e., the match between the user intent (expressed as a query)
    and the \emph{content} of the document.
    In this work we want to draw attention to two aspects
    of retrieval system performance affected by the \emph{presentation} of results:
    result attractiveness (``perceived relevance'') and immediate usefulness
    of the snippets (``snippet relevance'').
    Perceived relevance may influence discoverability of good topical documents
    and seemingly better rankings may in fact be less useful to the user
    if good-looking snippets lead to irrelevant documents or vice-versa.
    And result items on a search engine result page (SERP) with high snippet relevance may
    add towards the total utility gained by the user even without the need to click
    those items.

    We start by motivating the need to collect different aspects of
    relevance (topical, perceived and snippet relevances)
    and how these aspects can improve evaluation measures.
    We then discuss possible ways to collect these relevance aspects
    using crowdsourcing and the challenges arising from that.
\end{abstract}



\section{Introduction}
\label{sec:introduction}

For decades the main evaluation paradigm for search engines was the Cranfield
methodology~\cite{cleverdon1966aslib}.
In a typical setting of a TREC conference,
the documents are evaluated by human raters who assign relevance labels
based on their judgement about the relevance of the document
to the user's topic of interest, expressed as a query.
A graded relevance scale is typically used with topical
relevance labels ranging from $0$ to $4$ or from \emph{irrelevant} to \emph{highly relevant}.

These relevance labels can be obtained either from trained experts or using
a crowdsourcing approach. Either way, cases of disagreement have to be addressed,
and those are usually treated as raters' mistakes, but may also arise from different
interpretations of the user intent or the notion of relevance.
In a traditional evaluation approach a single relevance label is chosen
for each document-topic pair. These labels are then aggregated to SERP-level
quality measures such as DCG~\cite{jarvelin2002cumulated} or ERR~\cite{chapelle2009expected}.
By using \emph{additional} inputs from raters, we can
\begin{inparaenum}[(a)]
\item refine these quality measures and
\item better understand the performance of retrieval systems.
\end{inparaenum}


\section{Related Work}
\label{sec:related_work}

The idea to separate \emph{perceived} and \emph{topical} relevance was suggested
by~\cite{chapelle2009dynamic} while designing the DBN click model. Unlike earlier click models,
it suggests that the likelihood of a user clicking a document depends
not on the topical relevance of the document, but rather on its
perceived relevance, since the user can only judge based on the
result snippet.
This idea was later picked up by~\cite{turpin2009including} who showed
that while topical and perceived relevance are correlated,
there is a noticeable discrepancy between them.
They performed a simulated experiment by modeling the user click
probability and showed that taking it into account would lead
to substantially different ordering of the systems
participating in a TREC Web Track.

The idea to separate out \emph{snippet} relevance appears after
the introduction of good abandonment~\cite{li2009good}:
cases when users abandon a search result page without clicking any results
and yet they are satisfied.
This may be due to the SERP being rich with instant answers~\cite{chilton2011addressing},
e.g., a weather widget or a dictionary box,
or due to the fact that a query has a precise informational need,
that can easily be answered in a result snippet~\cite{chuklin2012good}.
In fact, as was shown by \cite{stamou2010interpreting} a big portion of abandoned searches
was due to a pre-determined behaviors: users came to a search engine
with a prior intention to find an answer directly on a SERP\@.
This is especially true when considering mobile search
where the internet connection can be slow
or the user interface is less convenient to use.
We complement these works by arguing
that good and relevant snippet does not necessarily lead to
a complete good abandonment, but rather represents an aspect of utility gained
by the user that is currently ignored.


\section{Application to Evaluation}
\label{sec:application_to_evaluation}
As was suggested by~\cite{carterette2011system}, many evaluation metrics,
including DCG and ERR may be viewed as based on a click model.
This was further refined by~\cite{chuklin2013click} where a recipe of
converting any click model into a metric was presented:
\begin{equation}
    \mathit{uMetric} = \sum_{k=1}^N P(C_k = 1) \cdot R_k \label{eq:umetric},
\end{equation}
where $R_k$ is the (topical) relevance of the $k$-th document in the ranking,
and $P(C_k = 1)$ is the probability that the user will click on that document.
Depending on the user model, the click probability may depend on
attractiveness parameters.
This is where we can use perceived relevance labels $A_k$ (attractiveness).
For example, for a metric based on the DCM model~\cite{guo2009efficient}
we have:\footnote{A similar but more involved equation can be obtained
for a metric based on the DBN model~\cite{chapelle2009dynamic}.}
\begin{equation}
    \mathit{uDCM} = \sum_{k=1}^N a(A_{k}) \prod_{i=1}^{k-1} \left(1 - a(A_{i}) s_i \right) \cdot R_k,
\end{equation}
where $a(A)$ is a list of parameters, one for each possible value
of perceived relevance label $A$; $s_i$ is another list of parameters,
one for each value of the document position $i$.

Further, if we want to use snippet relevance labels $S_k$, we introduce a metric
of the utility gained from the SERP itself similar to \eqref{eq:umetric}:
\begin{equation}
    \mathit{uMetric}_S = \sum_{k=1}^N P(E_k = 1) \cdot S_k \label{eq:umetricS},
\end{equation}
where $P(E_k = 1)$ is the probability that the user examines the $k$-th document.
Again, for DCM that would lead us to:
\begin{equation}
    \mathit{uDCM}_S = \sum_{k=1}^N \prod_{i=1}^{k-1} \left(1 - a(A_{i}) s_i \right) \cdot S_k.
\end{equation}
To summarize, we showed that by collecting perceived, topical and snippet relevance
we can refine system quality measures (eq.~\eqref{eq:umetric}, \eqref{eq:umetricS}).
To estimate the effect of this refinement one can compute correlations
with online click metrics similar to~\cite{chuklin2013click}
or with side-by-side comparison judgements collected using independent set of raters.


\section{Gathering Judgements}
\label{sec:gathering_relevance_judgements}
Now that we have argued that perceived, topical and snippet relevance are
potentially valuable dimensions of assessing system quality,
how do we gather the required judgements?
Firstly, we believe, that the \emph{topical} relevance definition
used by TREC raters is time-tested and hence can be used without modification.
Secondly, \emph{snippet} relevance can be treated as document topical relevance
with document replaced by its snippet.
We also need additional messaging for the raters explaining to them why the
``documents'' are so short to avoid undervalued scores.
In order to prevent the raters from confusing this task with perceived relevance judgement,
we may hide the fact that they are judging clickable snippets and just
refer to them as short summaries.\footnote{It would be interesting
    to alter this clause in the rater's instruction and see how the outcomes change.}
Similar ratings were collected by~\cite{chuklin2012good}, where three possible answers
were offered to the raters: the snippet ``answers the user question,''
``answers the question only partially,''
``does not answer the question.''
Third and finally, \emph{perceived} relevance is a new task that has to be formulated
by explaining to the rater the story of a web search and asking her if she
would click this link in order to find the relevant information in the document.
The snippet has to be shown without the context of the other snippets
and without its placement on a SERP to avoid position and presentation biases.

These, then, are the challenges of gathering relevance judgements in
the multi-aspect setting that we are proposing:
\squishlisttwo
    \item How to make sure the raters do not confuse different tasks (topical, snippet, perceived relevance)?
    \item How do we treat special SERP items such as images, instant answers or interactive tools?
    \item What influence does the query category have on the difficulty of the task?
        For example, snippet relevance does not make sense for navigational queries.
\squishend

\section{Conclusion}
This paper advocates for the need to review the notion of relevance
in to order improve evaluation as well as understand the anatomy of relevance.
We believe that after performing initial experiments and collecting feedback
from the raters, we can address the challenges outlined above
and derive a judgement procedure that will allow us
to collect all three aspects of relevance,
refine system performance evaluation and
get deeper insights into the foundation of relevance.


\smallskip
\begin{spacing}{1}
\smallskip\noindent\scriptsize{\bf Acknowledgements.}
This research was partially supported by
the European Community's Seventh Framework Programme (FP7/2007-2013) under
grant agreements nrs 288024 and 312827,
the Netherlands Organisation for Scientific Research (NWO)
under pro\-ject nrs
727.\-011.\-005, 
612.001.116, 
HOR-11-10, 
640.006.013, 
the Center for Creation, Content and Technology (CCCT),
the QuaMerdes project funded by the CLARIN-nl program,
the TROVe project funded by the CLARIAH program,
the Dutch national program COMMIT, 
the ESF Research Network Program ELIAS,
the Elite Network Shifts project funded by the Royal Dutch Academy of Sciences (KNAW),
the Netherlands eScience Center under project number 027.012.105,
the Yahoo! Faculty Research and Engagement Program,
the Microsoft Research PhD program,
and
the HPC Fund.
\end{spacing}

\bibliographystyle{abbrv}
\bibliography{gear2014-perceived}
\end{document}